\documentclass[letterpaper, 10 pt, conference]{ieeeconf}  

\IEEEoverridecommandlockouts                              
\usepackage{graphics} 
\usepackage{epsfig} 
\usepackage{mathptmx} 
\usepackage{times} 
\usepackage{amsmath} 
\usepackage{amssymb}  
\usepackage{color}
\usepackage{cite}
\usepackage{flushend}

\title{\LARGE \bf
Matrix Perturbation Theory of Inter-Area Oscillations
}

\author{J. Fritzsch, M. Tyloo, and Ph. Jacquod
\thanks{This work has been supported by the Swiss National Science Foundation under grant 200020\_182050}
\thanks{J. Fritzsch, M. Tyloo, and Ph. Jacquod are with the Department of Quantum Matter Physics, University of Geneva, CH-1211 Geneva, Switzerland and 
the School of Engineering, University of Applied Sciences of Western Switzerland HES-SO, CH-1951 Sion, Switzerland.}
\thanks{Emails:\tt{\small{julian.fritzsch@etu.unige.ch, \{melvyn.tyloo, philippe.jacquod\}@unige.ch}}}%
}
\begin{document}

\maketitle
\thispagestyle{empty}
\pagestyle{empty}

\begin{abstract}
Interconnecting power systems has a number of advantages such as better electric power quality, 
increased reliability of power supply, economies of scales through production and reserve pooling and so 
forth. Simultaneously, it may jeopardize the overall system stability with the emergence of so-called inter-area oscillations, which are coherent oscillations involving groups of rotating machines separated by large distances up to thousands of kilometers. These often weakly damped modes
may have harmful consequences for grid operation, yet despite decades of investigations, 
the mechanisms that generate them are still poorly understood, and the existing theories are based on 
assumptions that are not satisfied in real power grids where such modes are observed.
Here we construct a matrix perturbation theory of large interconnected power systems that clarifies
the origin and  the conditions for the emergence of inter-area oscillations. We show that coherent inter-area oscillations
 emerge from the zero-modes of a multi-area network Laplacian matrix, which
hybridize only weakly with other modes, even under significant capacity of the inter-area tie-lines, i.e. even
when the standard assumption of area partitioning is not satisfied. The general theory
is illustrated on a two-area system, and numerically applied to the well-connected PanTaGruEl model of the synchronous grid of continental Europe. 
        \end{abstract}

        \section{\textsc{Introduction}}
        Recent decades have witnessed a tendency to interconnect already large power transmission grids into
        larger and larger systems.
        Such interconnection is beneficial as it generally improves power quality, in particular voltage and frequency stability, it guarantees the safe and reliable supply of electric energy from the resulting
        diversification of power generation and it enables production and reserve pooling which 
        leads to economies of scales~\cite{Machowski2008,Rogers2012}. These advantages come however with negative side effects, perhaps the most important one being inter-area oscillations~\cite{Rogers2012,Klein1991}.
        These long-range modes  
        have been observed in continental transmission grids, where they manifest themselves as coherent oscillations of geographically separated groups of 
generators against each other~\cite{Entsoe2017,WECC}. 
With the ever increasing penetration of new renewables in power grids, and the associated reduction in overall inertia, there is a risk that these modes will occur more frequently. When present, these modes effectively reduce line capacities,
may damage rotating machines and, when not sufficiently damped as is often the case, may 
eventually trigger cascading failures and induce blackouts~\cite{Powertechlabs1997,Kosterev1999,Rogers2012,WECC,Entsoe2017}. It is therefore crucial to 
understand their origin, the conditions under which they occur and how to damp these modes. Below we address the first of these pressing issues. 
        
        The literature on inter-area oscillations is vast and here we mention only some typical works. A first direction of research is essentially phenomenological, where numerical modal analysis is used to investigate the damping of the slow modes of the swing equations~\cite{Janssens2000,Grebe2010,Rogers2012}. 
Another approach has been to construct such modes starting from network reduction algorithms into equivalent models~\cite{Cheng2021}, aggregating entire areas into single nodes~\cite{Kokotovic1976,Chow1982,Cho84,Cho85,Cho91}. The procedure assumes a separation of time scales between the intra- and inter-area dynamics, justifying a singular perturbation approximation~\cite{Kokotovic1976,Chow1982}. Strictly speaking, this assumption presupposes a partition of the network into areas with strong intra-area couplings and weak inter-area connectivity. The resulting mathematical conditions on the network structure that need to be fulfilled are 
never satisfied in real power networks, however. A rigorous understanding of inter-area oscillations in realistic settings is therefore still lacking.
        
In contrast to earlier works on inter-area oscillations, we start here from an a priori non-partitioned power network. Applying one of the many existing aggregation algorithms~\cite{Chow1982,Cheng2021}, we model the network as a collection of $r$ well connected areas. The number $r$ of areas is somewhat arbitrary, however it needs to be sufficiently larger than the number of inter-area oscillations one would like to construct. We introduce a parameter $\epsilon \in [0,1]$ multiplying the capacity of each tie-line between any two areas which allows us to interpolate between disconnected areas when $\epsilon=0$ and the original network when $\epsilon=1$. We apply matrix perturbation theory 
to this model, which gives Taylor expansions in $\epsilon$ for the eigenvalues and -vectors of the full network Laplacian matrix as a function of those of the disconnected Laplacian.  
We find that slow network modes corresponding to inter-area oscillations originate from the hybridization of the $r$ zero-modes of the disconnected Laplacian. Our second, main result is that, while matrix perturbation theory globally breaks down at small values of $\epsilon$ corresponding to the mathematical conditions justifying the standard singular perturbation approximation~\cite{Kokotovic1976,Chow1982}, our theory 
nevertheless remains locally justified upon restoration of the original network, $\epsilon \rightarrow 1$ for 
several of the slowest modes, which retain their structure as the inter-area tie-line are
restored with their original capacity. 
        
        The manuscript is organized as follows. Section~\ref{SecII} introduces our mathematical notations. Section~\ref{SecIII} gives the power network model we consider and its linearization around an operational state. Section~\ref{SecIV} gives a brief overview of matrix perturbation theory and gives eigenvalues and eigenvectors corrections. Of particular interest is the case of a matrix with repeated, i.e. degenerate eigenvalues. Section~\ref{SecV} applies matrix perturbation theory to inter-area oscillations. In particular it gives an upper bound on the inter-area connection strength for the validity of perturbation theory, and discusses avoided crossings that hamper the approximation. Section~\ref{SecVI} illustrates the theory first 
on a simple two-area network, then on a realistic model of the synchronous transmission grid of continental Europe. Conclusions are given in Section~\ref{SecVII}.
       
\section{\textsc{Mathematical Notation and Definitions}}\label{SecII}
We consider a network with $N$ nodes, which is subdivided into $r$ areas, each with $n_i$, $i=1, \ldots r$
nodes. We write column vectors $\mathbf{v} \in \mathbb{R}^n$ as bold lowercase letters and their transpose (row) vectors as $\mathbf{v}^\top$.
Matrices $\mathbf{M} \in \mathbb{R}^{n\times q}$ are written in bold uppercase letters.
Block diagonal matrices are written as $\mathbf{M} = \mathrm{diag}(\mathbf{M}_1, \mathbf{M}_2,\dots)$.
The $j^{\rm th}$ unit vector with one nonzero component is denoted $(\hat{e}^j)_i = \delta_{ij}$.
The vector of zeros (ones) of dimension $n$ is denoted $\mathbf{0}_n$ ($\mathbf{1}_n$).

\section{\textsc{Model}}\label{SecIII}
The transient and small-signal dynamics of high voltage AC power grids are commonly modeled by the swing equations. They  describe the dynamics of voltage angles $\theta_i$, assuming constant voltage amplitudes
$V_0$ over the whole network. 
At very high voltage, it is convenient to use the lossless line approximation which reads~\cite{Machowski2008,Bergen1981} 
\begin{subequations}\label{eq:swingequation}
\begin{align}
        m_i \dot{\omega_i} + d_i \omega_i &= p_i -\sum_j  B_{ij} V_0^2 \sin(\theta_i-\theta_j), \,\, i \in V_g,\label{eq:swingequationprod}\\
        d_i\omega_i &= p_i - \sum_j  B_{ij} V_0^2 \sin(\theta_i-\theta_j), \,\, i \in V_c.\label{eq:swingequationload}
\end{align}
\end{subequations}
These equations are written in a frame rotating at the rated grid frequency of $f = 50$ or $60\,\mathrm{Hz}$.
Each network node is either a generator ($i \in V_g$) or a consumer ($i \in V_c$),  
with  inertia $m_i$, damping, $d_i$, voltage angle $\theta_i$, frequency $\omega_i=\dot{\theta_i}$
and an active power, $p_i > 0$ for generators and $p_i < 0$ for consumers. 
In \eqref{eq:swingequationload}, 
consumers are modeled as frequency-dependent loads~\cite{Bergen1981}.
The nodes $i$ and $j$ are connected by a power line with susceptance $B_{ij}$.

Given the active power vector $\mathbf{p}^{(0)}$, 
the operational state $\pmb{\theta^{(0)}}$ is the stationary solution to \eqref{eq:swingequation}
satisfying, for each component of  $\mathbf{p}^{(0)}$,
\begin{equation}
        p_{i}^{(0)} = \sum_j  B_{ij}\, V_0^2 \, \sin(\theta_{i}^{(0)}-\theta_{j}^{(0)}) \, .
\end{equation}
For a small signal perturbation $\mathbf{p}^{(0)} \to \mathbf{p}^{(0)} + \pmb{\delta p}$, angle and frequency dynamics are governed by the linearized swing equations
\begin{equation}\label{eq:linswing}
        \mathbf{M}\dot{\pmb{\omega}} + \mathbf{D}\pmb{\omega} = \pmb{\delta p} - \mathbf{L}\pmb{\delta\theta} \, , 
\end{equation}
where 
$\mathbf{M} = \mathrm{diag}({m_i})$ [$m_i = 0$ for consumers], $\mathbf{D}=\mathrm{diag}({d_i})$, $\pmb{\delta\theta} = \pmb{\theta} - \pmb{\theta^{(0)}}$, $\pmb{\omega} = {\pmb{\delta \dot\theta}}$, and
\begin{equation}
        \mathbf{L}_{ij} = \begin{cases}
                -B_{ij}\, V_0^2 \, \cos(\theta^{(0)}_i-\theta^{(0)}_j) &\text{for } i\neq j \\
                \sum_k B_{ik}\, V_0^2 \,\cos(\theta^{(0)}_i-\theta^{(0)}_k) &\text{for } i = j \, , 
        \end{cases}
\end{equation}
is the weighted network Laplacian matrix. When the network is singly-connected, it is
positive semidefinite with one zero eigenvalue,  $\lambda_1 = 0$, corresponding to the 
eigenvector $\mathbf{u}_1 = (1,1,\dots,1)^\top/\sqrt{N}$.
When the network consists of $r$ disconnected areas, the Laplacian matrix has $r$ zero eigenvalues, each of which corresponding to an eigenvector with constant components within an area. Following the standard physics denomination, we call such eigenvectors ``zero-modes''.

\section{\textsc{Matrix Perturbation Theory}}\label{SecIV}

Matrix perturbation theory addresses the question ``can we express the eigenvalues and -vectors of a matrix, from the known eigenvalues and -vectors of a slightly different matrix'' The answer is yes. As long as the difference between the two matrices is not too large, the eigenvalues and -vectors of the ``perturbed'' matrix are expressed in a controlled series expansion in the eigenvalues and -vectors of the ``unperturbed'' matrix.
The method is well-known and widely used in physics, and it has recently been applied to problems
in electric power systems~\cite{Yang2018,Pagnier2019b, Coletta2020,Bamieh2020}. Situations where 
the unperturbed matrix has a non-degenerate spectrum (where all its eigenvalues are distinct) must be 
treated differently from cases with a degenerate spectrum (where some of the eigenvalues are repeated).
For the sake of completeness we present a short introduction to matrix perturbation theory following~\cite{J.J.Sakurai2017}.

\subsection{Non-Degenerate Case}

Consider that
the Laplacian matrix $\mathbf{L}$ is 
\begin{equation}\label{eq:laplacianpartition}
        \mathbf{L} = \mathbf{L}_0 + \epsilon \mathbf{L}_I,
\end{equation}
where $\epsilon \in [0, 1]$ scales the strength of the perturbation. We call $\mathbf{L}_0$ the unperturbed Laplacian  and $\epsilon \mathbf{L}_I$ the perturbation. 
One wants to give a controlled approximation to the full eigenvalue problem
\begin{equation}
        (\mathbf{L}_0 + \epsilon\mathbf{L}_I) \mathbf{u}_\alpha = \lambda_\alpha \mathbf{u}_\alpha.\label{eq:fullproblem}
\end{equation}
from the known eigenvalues and -vectors ($\lambda_\alpha^{(0)}, \mathbf{u}_\alpha^{(0)}$)
of the unperturbed problem, $\epsilon=0$.
The trick is to expand $\mathbf{u}_\alpha$ and $\lambda_\alpha$ as
\begin{subequations}
        \label{eq:seriesex}
        \begin{align}
                \mathbf{u}_\alpha &= \mathbf{u}_\alpha^{(0)} + \epsilon \mathbf{u}_\alpha^{(1)} + \epsilon^2 \mathbf{u}_\alpha^{(2)} + \dots,\\
                \lambda_\alpha &= \lambda_\alpha^{(0)} + \epsilon \lambda_\alpha^{(1)} + \epsilon^2 \lambda_\alpha^{(2)} + \dots.
        \end{align}
\end{subequations}
The coefficients in these expansions 
can be obtained order by order~\cite{Bamieh2020,J.J.Sakurai2017}, and
in this paper we will restrict ourselves to the first order corrections to the eigenvectors
\begin{equation}
        \mathbf{u}_\alpha^{(1)} = \sum_{\beta\neq\alpha}\frac{{\mathbf{u}_\beta^{(0)^\top}} \, \mathbf{L}_I \, \mathbf{u}_\alpha^{(0)}}{\lambda_\alpha^{(0)}-\lambda_\beta^{(0)}}\mathbf{u}_\beta^{(0)}\label{eq:firstorderveccorr}
\end{equation}
and the first and second order  corrections to the eigenvalues
\begin{subequations}
\begin{align}
        \lambda_\alpha^{(1)} &= \mathbf{u}_\alpha^{(0)^\top}\mathbf{L}_I\mathbf{u}_\alpha^{(0)},\label{eq:firstordervalcorr}\\
        \lambda_\alpha^{(2)} &
       = \sum_{c}\frac{\Big|\mathbf{u}_\alpha^{(0)^\top}\mathbf{L}_I\mathbf{u}_\beta^{(0)}\Big|^2}{\lambda_\alpha^{(0)}-\lambda_\beta^{(0)}}.\label{eq:secordervalcorr}
\end{align}
\end{subequations}
Higher-order terms have similar, though more complicated structures and we do not discuss them here. Suffice it to mention that, from \eqref{eq:firstorderveccorr} and \eqref{eq:secordervalcorr},  the convergence of the perturbation expansions \eqref{eq:seriesex} for all $\alpha$'s requires that
\begin{equation}\label{eq:globalcon}
        \Big|\lambda_\alpha^{(0)} - \lambda_\beta^{(0)}\Big| \gg \epsilon \Big|{\mathbf{u}_\beta^{(0)^\top}}\mathbf{L}_I \mathbf{u}_\alpha^{(0)}\Big| \, ,\,\,\, \,\,\, \forall \alpha,\beta \, .
\end{equation}
Below we show that this condition is equivalent to the standard condition for the validity of singular 
perturbation theory~\cite{Chow1982, Chow2013}. 
One of our main findings will be however that matrix perturbation theory breaks down later for slow
modes with small $\lambda_\alpha^{(0)}$, so that several inter-area modes can be constructed for larger $\epsilon \rightarrow 1$ 
in large, well-connected networks.

\subsection{Degenerate Case}\label{chapter:deg}

Clearly, \eqref{eq:firstorderveccorr} and \eqref{eq:secordervalcorr} exhibit divergences if two (or more) eigenvalues
are equal. Therefore matrix perturbation theory treats the degenerate case differently.

One considers separately each degenerate subspace $\mathcal{L}_a = \mathrm{span}(\{\mathbf{u}_\alpha^{(0)}\}: \mathbf{L}_0\mathbf{u}_\alpha^{(0)} = \lambda_\alpha^{(0)}\mathbf{u}_\alpha^{(0)}, \lambda_\alpha^{(0)} = \lambda_a)$
corresponding to each multiply-repeated eigenvalue $\lambda_a$ of $\mathbf{L}_0$. The set 
$\{\mathbf{u}_\alpha^{(0)}\}$ of $r_a$ degenerate eigenvectors
gives an orthonormal basis of $\mathcal{L}_a$, so that 
approximate eigenvectors of $\mathbf{L}_0 + \epsilon \mathbf{L}_I$ within $\mathcal{L}_a$ are given by
linear combinations of these eigenvectors, with coefficients given by the column of the orthogonal matrix $\mathbf{O}$
diagonalizing the projection $\mathbf{V}_a$ of $\mathbf{L}_I$ onto $\mathcal{L}_a$,
\begin{equation}\label{eq:Vdiag}
       \mathbf{V}_a=\mathbf{O}^\top \mathrm{diag}({\lambda_\alpha}^{(1)}) \, \mathbf{O} \, .
\end{equation}
Here the $r_a \times r_a$ matrix $\mathbf{V}$ has elements 
\begin{equation}
       (\mathbf{V}_a)_{\alpha,\beta} = \mathbf{u}_\alpha^{(0)^\top} \mathbf{L}_I \mathbf{u}_{\beta}^{(0)}  \, .
\end{equation}
From \eqref{eq:Vdiag}, the first order corrections to each degenerate eigenvalue $\lambda_\alpha^{(0)}=\lambda_a$ are given
by $\epsilon \lambda_\alpha^{(1)}$, with the eigenvalues $\lambda_\alpha^{(1)}$ of the reduced interaction matrix $\mathbf{V}_a$. The eigenvectors of the 
latter also determine the relevant linear combination of degenerate eigenvectors in $\mathcal{L}_a$.
As long as $\epsilon$ is sufficiently small, these give the dominant corrections to the degenerate part of the 
spectrum of $\mathbf{L}_0$.

Once $\lambda_\alpha^{(0)} + \epsilon \lambda_\alpha^{(1)}$ approaches the part of the spectrum outside
$\mathcal{L}_a$, second order corrections are no longer negligible. They
are given by~\eqref{eq:secordervalcorr} with the sum over $\beta$ 
being over eigenvalues and -vectors outside of
$\mathcal{L}_a$.

\section{\textsc{From Zero-Modes to Inter-Area Oscillations}}\label{SecV}

We next apply matrix perturbation theory to construct the slow modes of a large, well-connected
Laplacian matrix $\mathbf{L}$, corresponding to inter-area oscillations in the power grid modeled
by  $\mathbf{L}$. The first step is to subdivide the system into $r$ areas using one of the existing
algorithms to do so~\cite{Chow1982,Cheng2021}. Unless the considered grid is originally partitioned into 
weakly connected areas, this subdivision is arbitrary and a priori not justified, but we will see how it enables
to construct slow modes of $\mathbf{L}$ when $r$ is chosen appropriately. We write 
\begin{equation}\label{eq:laplacianpartitionnoe}
        \mathbf{L} = \mathbf{L}_0 + \mathbf{L}_I,
\end{equation}
with 
\begin{equation}\label{eq:l0partitioning}
        \mathbf{L}_0 = \mathrm{diag}\left(\mathbf{L}_0^{(1)},\dots,\mathbf{L}_0^{(r)}\right),
\end{equation}
where $\mathbf{L}_0^{(i)}$ denotes the internal Laplacian of the $i^{\rm th}$ area.
To apply matrix perturbation theory, we consider \eqref{eq:laplacianpartition} instead of
\eqref{eq:laplacianpartitionnoe}, keeping in mind that in the end, we need to take the limit $\epsilon \rightarrow 1$. 

The unperturbed Laplacian $\mathbf{L}_0$ has $r$ zero eigenvalues.
The corresponding eigenvectors have constant components in each area and can be any linear combination
of the area zero-modes
\begin{equation}\label{eq:zeromodes}
        \mathbf{v}_i = \frac{1}{\sqrt{n_i}}(0,\dots,0,\mathbf{1}_{n_i},0,\dots,0)^\top \, , \,\,\, i=1, \ldots r \, .
\end{equation}
The perturbation $\mathbf{L}_I$ is also Laplacian and contains the inter-area connections.
Increasing $\epsilon$ in \eqref{eq:laplacianpartition} changes the network from a disconnected one 
into the original, fully connected network.
We apply degenerate perturbation theory to the zero-modes. We will see that 
slow, inter-area modes arise from the hybridization of some of the area zero-modes in \eqref{eq:zeromodes}.


Our first step is to write $\mathbf{L}$ in a basis that diagonalizes $\mathbf{L}_0$,
\begin{align}
        \begin{split}
        \mathbf{U}^\top\mathbf{L}\mathbf{U} = \mathbf{U}^\top(\mathbf{L}_0+\epsilon\mathbf{L}_I)\mathbf{U}=\widetilde{\mathbf{L}}_0 + \epsilon\widetilde{\mathbf{L}}_I \, , 
\end{split}
\end{align}
by means of the $N \times N$ matrix 
\begin{equation}
        \mathbf{U} = (\mathbf{u}_1^{(0)}, \dots, \mathbf{u}_N^{(0)}),
\end{equation}
whose columns contain the components of the eigenvectors of $\mathbf{L}_0$. Because the latter
has $r$ degenerate zero-modes, the $r$ first columns of $\mathbf{U}$ 
can be chosen as arbitrary linear combinations
of the area zero-modes in \eqref{eq:zeromodes}. We chose the combination that diagonalizes 
$\mathbf{L}_I$ in the degenerate subspace $\mathcal{L}_0$, $\mathbf{L}_I \mathbf{u}_\alpha^{(0)}
= \lambda_\alpha^{(1)} \mathbf{u}_\alpha^{(0)}$, $\alpha=1,2, \ldots r$. It is straightforward to see then that 
\begin{align}
        \widetilde{\mathbf{L}}_0  &= \mathrm{diag}\left(\mathbf{U}^\top(\mathbf{L}_0+\epsilon\mathbf{L}_I)\mathbf{U}\right) \, ,\\
        \widetilde{\mathbf{L}}_I& =  \mathbf{U}^\top \, \mathbf{L}_I \, \mathbf{U} - \mathrm{diag}\left(\mathbf{U}^\top \, \mathbf{L}_I \, \mathbf{U}\right) \, .
\end{align}
Spectral corrections up to the first order in $\epsilon$ are contained in 
$\widetilde{\mathbf{L}}_0$ while higher-order corrections come from  
$\epsilon \widetilde{\mathbf{L}}_I$.

We consider $\widetilde{\mathbf{L}}_0$ as the unperturbed matrix and $\epsilon \widetilde{\mathbf{L}}_I$ as the perturbation matrix. Because $\widetilde{\mathbf{L}}_0$ is diagonal in the basis we use, 
the unperturbed eigenvectors have components $\tilde{u}_{\alpha,j}^{(0)} = \delta_{\alpha j}$.
From~\eqref{eq:firstordervalcorr} and~\eqref{eq:secordervalcorr} it directly follows that the first order corrections to the eigenvalues are zero, because we already included them in our definition of  $\widetilde{\mathbf{L}}_0$,
and that the second order corrections read
\begin{equation}\label{eq:secondorder}
        \lambda_\alpha^{(2)}  = \sum_{\beta\neq\alpha}\frac{\Big|\mathbf{u}_\alpha^{(0)^\top}\mathbf{L}_I\mathbf{u}_\beta^{(0)}\Big|^2}{(\widetilde{\mathbf{L}}_0)_{\alpha\alpha}-(\widetilde{\mathbf{L}}_0)_{\beta\beta}} \, .
\end{equation}
The correction to the zero-modes is given by \eqref{eq:secondorder} 
where $\alpha$ is one of the zero modes, $\beta$ is one of the non-zero modes. In that case,
the denominator may be written as
\begin{equation}
        (\widetilde{\mathbf{L}}_0)_{\alpha\alpha}-(\widetilde{\mathbf{L}}_0)_{\beta\beta} = -\lambda_\beta^{(0)} + \epsilon\big(\mathbf{u}_\alpha^{(0)^\top}\mathbf{L}_I\mathbf{u}_\alpha^{(0)}-{\mathbf{u}_\beta^{(0)}}^\top\mathbf{L}_I\mathbf{u}_\beta^{(0)}\big).
\end{equation}
This defines a critical value of $\epsilon$
below which there will be no divergence to any order in perturbation theory for the zero-mode $\alpha$,
\begin{equation}\label{eq:epsilonc}
        \epsilon_{c,\alpha} = \min_{\alpha, \epsilon_c \geq 0}\left(\frac{\lambda_\beta^{(0)}}{\mathbf{u}_\alpha^{(0)^\top}\mathbf{L}_I\mathbf{u}_\alpha^{(0)}-\mathbf{u}_\beta^{(0)\top}\mathbf{L}_I\mathbf{u}_\beta^{(0)}}\right) \, ,
\end{equation}
because for $\epsilon < \epsilon_{c, \alpha}$, there is no vanishing denominator in the perturbation expansion
for $\lambda_\alpha$.
Here $\min_{\alpha, \epsilon_c \geq 0}$ denotes the minimum over all $\beta$ that satisfy $\epsilon \geq 0$.
The criterion $\epsilon < \epsilon_{c,\alpha}$ is based on the distance between the zero-modes and 
nearby non-zero-modes and the slope of their variation with $\epsilon$ to leading order in perturbation theory. When the slope difference is small and the distance is large, our perturbative construction of 
zero-modes may remain justified beyond $\epsilon>1$, for areas connected even more strongly than in the
real network. Below we will see that this is the case for several modes in well connected, large networks. 

Two important remarks are in order here before we apply our theory to power grid models. First,
electric power grids are complex networks with no particular symmetry. Because of 
the absence of symmetries there are generically no degeneracies, except those of 
the zero-modes which are due to the Laplacian nature of the network coupling in each area. Second, upon
increasing $\epsilon$ from zero, eigenvalues of $\mathbf{L} = \mathbf{L}_0 + \epsilon \mathbf{L}_I$
move, first quasi-linearly up or down with $\epsilon$ - corresponding to the first-order corrections \eqref{eq:firstordervalcorr} -
before higher-order corrections kick in. The latter have an important consequence that, unless some 
symmetries are at work, eigenvalues may come very close to one another but they eventually repel each other and do not cross~\cite{Neumann1929}. It is in the immediate vicinity of the resulting avoided crossings
that eigenvectors exchange their structure. Conversely, eigenvectors corresponding to  
eigenvalues that do not undergo any avoided crossing as $\epsilon$ is varied 
do not change their structure by much. 
The threshold value $\epsilon_{c,\alpha}$ given in \eqref{eq:epsilonc} gives a parametric estimate
for the first location of an avoided crossing involving the $\alpha^{\rm th}$ zero-mode.
Below we show that low-lying zero-modes 
are not subjected to avoided crossings, protected as they are from the rest of the spectrum by higher-lying zero-modes. 
\section{\textsc{Inter-Area Oscillations in Power Grid Models}}\label{SecVI}

\subsection{Two-area Network}
We start by applying matrix perturbation theory to a simple two-area network.
The areas have $n_1$ and $n_2$ nodes respectively, and the unperturbed Laplacian is $\mathbf{L}_0 = \mathrm{diag}(\mathbf{L}_0^{(1)},\mathbf{L}_0^{(2)})$. 
It has two zero eigenvalues, corresponding to the area zero-modes, i.e. instead of \eqref{eq:zeromodes} 
one has
\begin{equation}
        \mathbf{v}_1 = (\mathbf{1}_{n_1}, \mathbf{0}_{n_2})^\top /\sqrt{n_1},\quad \mathbf{v}_2 = (\mathbf{0}_{n_1}, \mathbf{1}_{n_2})^\top /\sqrt{n_2}.
\end{equation}
The inter-area connections are captured by the interaction Laplacian $\mathbf{L}_I$.
The reduced interaction matrix $\mathbf{V}$ is then given by
\begin{equation}
        \mathbf{V} = 
        A \begin{pmatrix}
                1/n_1 & -1/\sqrt{n_1n_2}\\
                -1/\sqrt{n_1n_2} & 1/n_2
        \end{pmatrix},
\end{equation}
where $A = \mathrm{tr}(\mathbf{L}_I)/2$ is the sum of the capacities of the inter-area tie-lines.
The eigenvalues of $\mathbf{V}$ are
\begin{equation}
        \lambda_1 = 0, \qquad \lambda_2 = A \left( \frac{n_1 + n_2}{n_1n_2} \right).
\end{equation}
The appropriate linear combination of the area zero-mode is
\begin{align}
        \mathbf{{u}}_1^{(0)} &= \sqrt{\frac{n_1}{n_1+n_2}}\mathbf{v}_1 + \sqrt{\frac{n_2}{n_1+n_2}}\mathbf{v}_2 = \mathbf{1}_{n_1+n_2}/\sqrt{n_1 + n_2}\\
        \mathbf{{u}}_2^{(0)} &= \sqrt{\frac{n_2}{n_1 + n_2}} \mathbf{v}_1 - \sqrt{\frac{n_1}{n_1 + n_2}} \mathbf{v}_2.
\end{align}
These eigenvectors are the constant Laplacian mode and the well-known Fiedler mode~\cite{MiroslavFiedler1989}.
It can be shown that the constant mode remains unchanged at 
every order of perturbation theory, because $\mathbf{L}_I$ is Laplacian, however the second mode
will in general change with $\epsilon$.

In this simple two-area case, the analytic threshold \eqref{eq:globalcon} for the validity of the theory
can be estimated by using the known bound on 
the smallest non-zero eigenvalue in each area $\lambda_2^{(i)} < 4 \chi_i / (n_i D_i)$, where 
$\chi_i = \min_{l,m} |(\mathbf{L}_0^{(i)})_{lm}|$ and $D_i$ denotes the diameter of area $i$~\cite{Mohar1991}.
We then find the threshold value for the inter-area connections
\begin{equation}
        A < \left( \frac{n_1n_2}{n_1 + n_2} \right)4 \min_i\left(\frac{\chi_i}{n_i D_i} \right).
\end{equation}
This means for global convergence the areas should show strong intra-area connections and weak inter-area connections, in agreement with the standard criteria of \cite{Chow1982}.

\subsection{East-West Oscillations in the European Grid}

\begin{figure}[htbp]
        \centering
        \includegraphics[width=\columnwidth]{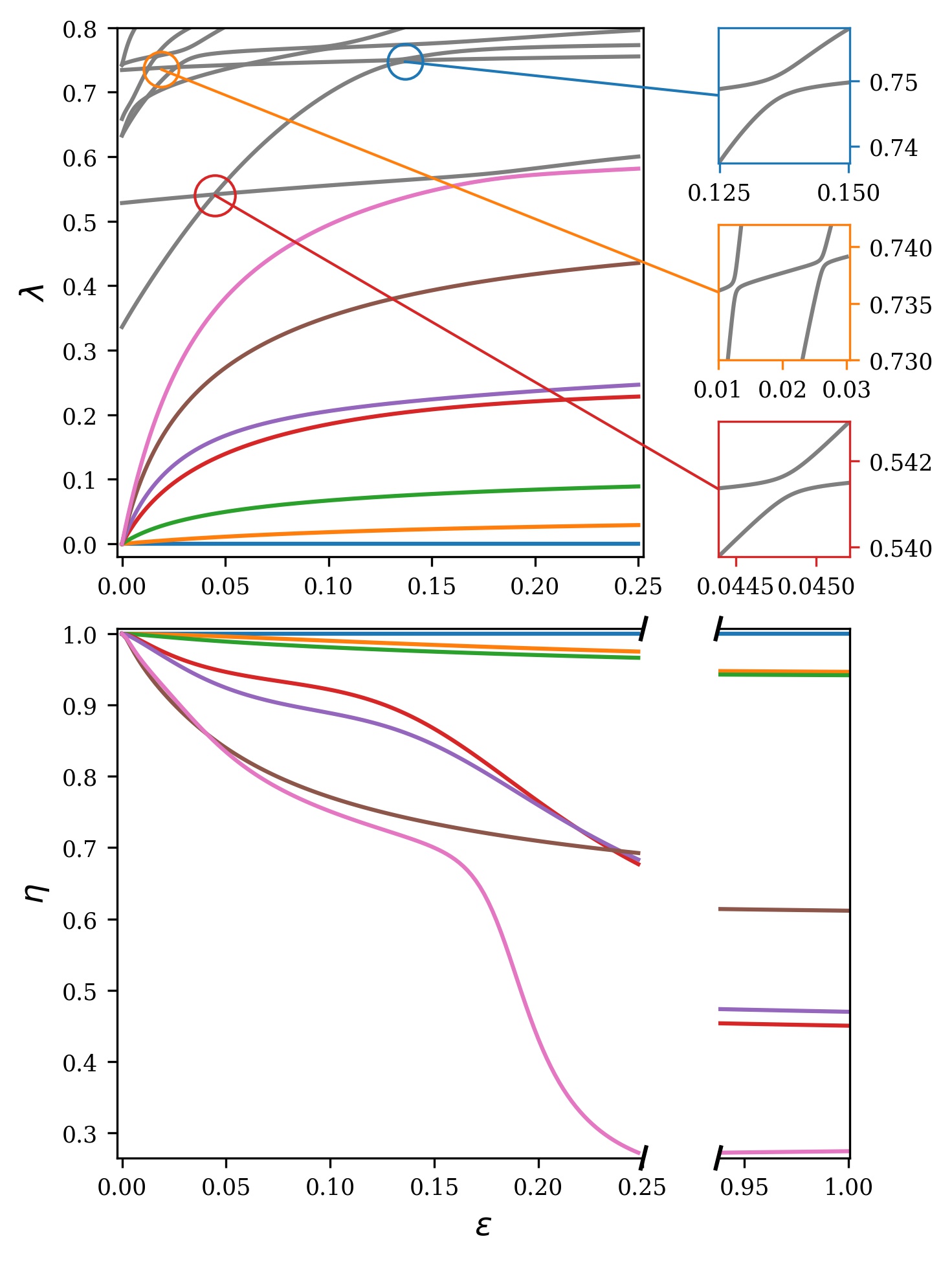}
        \caption{Top panels: Evolution of the eigenvalues of the Laplacian of \eqref{eq:laplacianpartition} and \eqref{eq:l0partitioning} with $r=7$ area partitioning of the
\emph{PanTaGruEl} model of the synchronous grid of continental Europe.
Zero-modes giving rise to inter-area oscillations are shown in color and some of the lowest non-degenerate modes in gray. The circles mark three illustrative avoided crossings. The three right panels make it clear that levels avoid crossing each other, because of the lack of specific symmetry in the system.
Bottom panel: evolution of the scalar product 
$\eta = \mathbf{u}_\alpha^\top(\epsilon=0) \cdot \mathbf{u}_\alpha(\epsilon)$ between hybridized zero-modes at $\epsilon=0$ and $\epsilon$. The avoided crossing at around $\epsilon=0.2$ between the seventh (pink) mode and the first non-degenerate (grey) leads to an abrupt drop in $\eta$ for the pink mode. 
Almost simultaneously, there is an avoided crossing between the fourth (red) and fifth (purple) eigenvalue
giving a noticeable drop in $\eta$ for both modes. The first three modes barely change their structure all the
way up to $\epsilon=1$.}
        \label{fig:avoidedcrossings7areas}
\end{figure}

We perform our main numerical investigations on the  \emph{PanTaGruEl}  model of the 
synchronous grid of continental Europe. The model is described in \cite{Pagnier2019,Tyloo2019}. It
consists of 3809 buses, 468 of which are generators, connected by 4944 power lines.
We first aggregate the model into seven areas using the algorithm of \cite{Chow1982, Chow2013},
and verify that all areas are connected.
In the top panel of Fig.~\ref{fig:avoidedcrossings7areas} the hybridization of the zero-modes and the evolution of some
of the lowest non-degenerate eigenvalues is shown as a function of $\epsilon \in [0.0, 0.25]$, 
 for a partition of seven areas. There are already several avoided crossings 
 at small $\epsilon$, indicating the overall breakdown of matrix perturbation theory, however these avoided
 crossing occur rather high in the spectrum and do not affect the lowest-lying zero-modes.
 This is confirmed by the mode-dependent threshold ~\eqref{eq:epsilonc} which is $\epsilon_{c,\alpha}>1$
 for $\alpha=2$ (the orange mode). Other modes have lower values, going down to  
 $\epsilon_{c,\alpha}\approx0.03$ for $\alpha=7$ (the pink mode) which indeed undergoes an avoiding crossing with higher-lying non-degenerate modes at $\epsilon \lesssim 0.18$.

The hybridization of the zero-modes can be quantified via the scalar product
\begin{equation}
\eta = \mathbf{u}_\alpha^\top(\epsilon=0) \cdot \mathbf{u}_\alpha(\epsilon) \, , 
\end{equation}
between an unperturbed mode at $\epsilon=0$ and its vector at $\epsilon \ne0$. The bottom panel 
of Fig.~\ref{fig:avoidedcrossings7areas} shows that the low-lying zero-mode essentially keep their
unperturbed structure, with 
$\eta \gtrsim 0.95$ all the way to the fully connected network limit $\epsilon \rightarrow 1$. Accordingly, 
the degenerate matrix perturbation theory presented above predicts the structure of the corresponding eigenvectors very well. These modes correspond to east-west inter-area oscillations~\cite{Breulmann2000}.

\begin{figure}[htbp]
        \centering
        \includegraphics[width=\columnwidth]{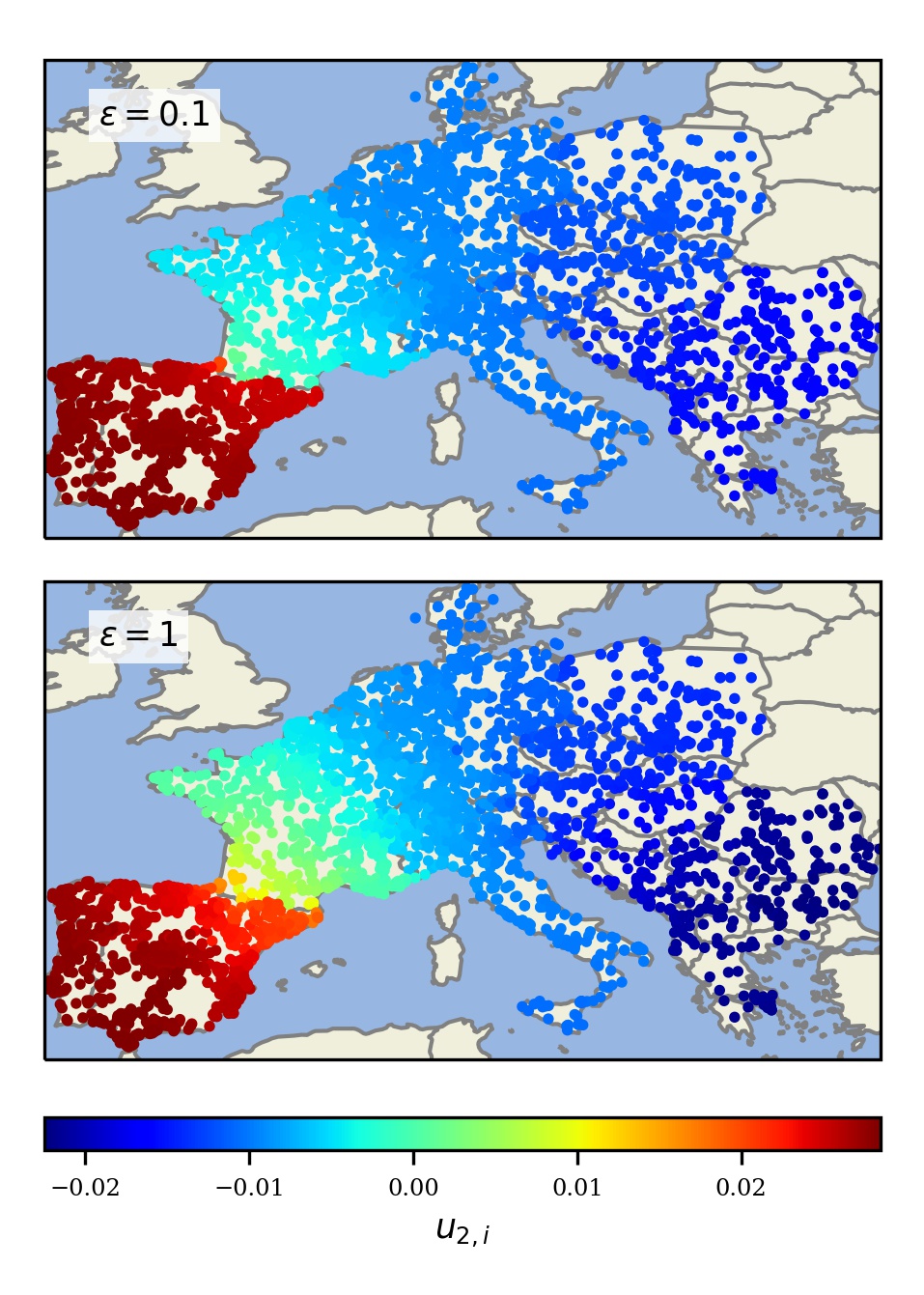}
        \caption{Structure of the Fiedler mode of the Laplacian of the \emph{PanTaGruEl} model of the synchronous grid of continental Europe in the weakly (top panel) and fully (bottom) connected cases. The color corresponds to the value of the eigenvector $u_{2,i}$ of the corresponding node $i$. The mode structure is essentially the same for $\epsilon=0.1$ as for $\epsilon=1$. This mode corresponds to east-west inter-area oscillations.}
        \label{fig:eigvecspanta}
\end{figure}

\begin{figure}[htbp]
        \centering
        \includegraphics[width=\columnwidth]{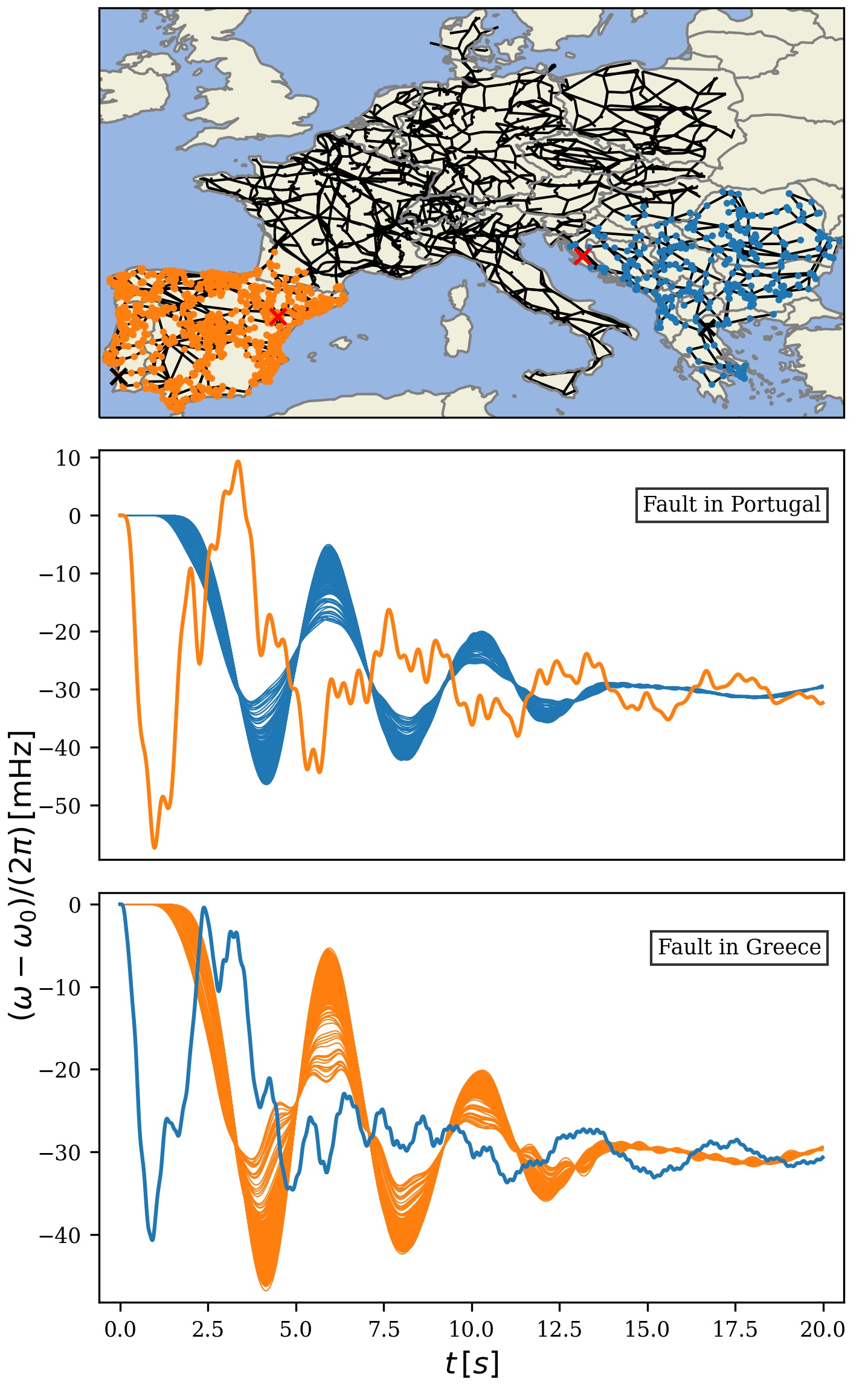}
        \caption{Time-evolution of frequencies following a 900~MW fault in Portugal (middle panel) or Greece (lower panel). The fault locations are marked by the two black crosses in the top panel. The response of all nodes in the area opposite to the fault, as well as on one node in the faulted area (red crosses in the top panel) are shown in the middle and lower panels.}
        \label{fig:timeevolution}
\end{figure}

We have observed, and will discuss in a follow-up paper that increasing 
the number $r$ of areas  improves the precision with which the lowest eigenvalues and -vectors
are predicted. Simultaneously, this increases the number of fast growing, initially degenerate eigenvalues,
which accordingly meet the non-degenerate eigenvalues at lower values of $\epsilon$. Because they
cannot cross them, however, they undergo avoided crossings which pushes back the initially 
non-degenerate part of the spectrum.  Qualitatively, this leads to a better protection of the low-lying 
eigenvalues which hybridize very little. This is further illustrated in   
Fig.~\ref{fig:eigvecspanta} which shows that the Fiedler, $\alpha=2$ mode 
keeps the same structure from the weakly coupled limit at
 $\epsilon=0.1$ to the fully coupled limit at $\epsilon=1$.
To show that these modes are indeed coupled to inter-area oscillations in the grid we 
finally investigate abrupt 900~MW power generation faults in Portugal and in Greece. The results are shown 
in Fig.~\ref{fig:timeevolution}. It is seen that
 following such a fault, the opposite area oscillates coherently, with all nodes oscillating at the same 
 phase and frequency, before synchronizing at a smaller frequency value.

\section{\textsc{Conclusions}}\label{SecVII}
We have constructed a matrix perturbation theory of slow coherent, inter-area oscillations. 
We have shown that inter-area oscillations emerge from the weak hybridization of some of the zero modes of coherent areas. The hybridization means that areas, even located far away from each other, are connected through a mode that is largely constant on each area. When such modes are excited, the corresponding 
areas oscillate coherently against each other. 
We finally stress that the singular perturbation theory of \cite{Chow1982, Chow2013} is absolutely 
not justified here, because the required small parameters have values $d=140$ and $\delta=37$, much larger than one. 
Work in progress investigates damping of inter-area oscillations and location of fault that might trigger them.

\section*{ACKNOWLEDGMENTS}
 We thank F.~D\"{o}rfler for discussions.




\begin{thebibliography}{10}
        \providecommand{\url}[1]{#1}
        \csname url@samestyle\endcsname
        \providecommand{\newblock}{\relax}
        \providecommand{\bibinfo}[2]{#2}
        \providecommand{\BIBentrySTDinterwordspacing}{\spaceskip=0pt\relax}
        \providecommand{\BIBentryALTinterwordstretchfactor}{4}
        \providecommand{\BIBentryALTinterwordspacing}{\spaceskip=\fontdimen2\font plus
        \BIBentryALTinterwordstretchfactor\fontdimen3\font minus
          \fontdimen4\font\relax}
        \providecommand{\BIBforeignlanguage}[2]{{%
        \expandafter\ifx\csname l@#1\endcsname\relax
        \typeout{** WARNING: IEEEtran.bst: No hyphenation pattern has been}%
        \typeout{** loaded for the language `#1'. Using the pattern for}%
        \typeout{** the default language instead.}%
        \else
        \language=\csname l@#1\endcsname
        \fi
        #2}}
        \providecommand{\BIBdecl}{\relax}
        \BIBdecl
        
        \bibitem{Machowski2008}
        J.~Machowski, J.~W. Bialek, and J.~R. Bumby, \emph{Power system dynamics:
          stability and control}, 2nd~ed.\hskip 1em plus 0.5em minus 0.4em\relax
          Chichester, U.K: Wiley, 2008.
        
        \bibitem{Rogers2012}
        G.~Rogers, \emph{Power system oscillations}.\hskip 1em plus 0.5em minus
          0.4em\relax Springer Science \& Business Media, 2012.
        
        \bibitem{Klein1991}
        M.~Klein, G.~J. Rogers, and P.~Kundur, ``A fundamental study of inter-area
          oscillations in power systems,'' \emph{IEEE Transactions on power systems},
          vol.~6, no.~3, pp. 914--921, 1991.
        
        \bibitem{Entsoe2017}
        Entsoe, ``Analysis of ce inter-area oscillations of 1st december 2016,''
          \emph{available on-line.}, 2017.
        
        \bibitem{WECC}
        WECC, ``Modes of inter-area power oscillations in western interconnection,''
          \emph{available on-line.}, 2013.
        
        \bibitem{Powertechlabs1997}
        {EPRI Final Report TR-108256}, ``System disturbance stability studies for
          western system coordinating council (wscc),'' \emph{prepared by Powertech
          Labs Inc.}, 1997.
        
        \bibitem{Kosterev1999}
        D.~N. {Kosterev}, C.~W. {Taylor}, and W.~A. {Mittelstadt}, ``Model validation
          for the august 10, 1996 wscc system outage,'' \emph{IEEE Transactions on
          Power Systems}, vol.~14, no.~3, pp. 967--979, 1999.
        
        \bibitem{Janssens2000}
        N.~Janssens and A.~Kamagate, ``Interarea oscillations in power systems,''
          \emph{IFAC Proceedings Volumes}, vol.~33, no.~5, pp. 217--226, 2000, iFAC
          Symposium on Power Plants and Power Systems Control 2000, Brussels, Belgium,
          26-29 April 2000.
        
        \bibitem{Grebe2010}
        E.~{Grebe}, J.~{Kabouris}, S.~{López Barba}, W.~{Sattinger}, and W.~{Winter},
          ``Low frequency oscillations in the interconnected system of continental
          europe,'' in \emph{IEEE PES General Meeting}, 2010, pp. 1--7.
        
        \bibitem{Cheng2021}
        \BIBentryALTinterwordspacing
        X.~Cheng and J.~Scherpen, ``Model reduction methods for complex network
          systems,'' \emph{Annual Review of Control, Robotics, and Autonomous Systems},
          vol.~4, no.~1, p. null, 2021. [Online]. Available:
          \url{https://doi.org/10.1146/annurev-control-061820-083817}
        \BIBentrySTDinterwordspacing
        
        \bibitem{Kokotovic1976}
        P.~V. Kokotovic, R.~E. O'Malley~Jr, and P.~Sannuti, ``Singular perturbations
          and order reduction in control theory—an overview,'' \emph{Automatica},
          vol.~12, no.~2, pp. 123--132, 1976.
        
        \bibitem{Chow1982}
        J.~H. Chow, Ed., \emph{{Time-Scale Modeling of Dynamic Networks with
          Applications to Power Systems}}.\hskip 1em plus 0.5em minus 0.4em\relax
          Berlin Heidelberg, Germany: Springer-Verlag, 1982.
        
        \bibitem{Cho84}
        J.~H. Chow, J.~Cullum, and R.~A. Willoughby, ``{A Sparsity-Based Technique for
          Identifying Slow-Coherent Areas in Large Power Systems},'' \emph{IEEE
          Transactions on Power Apparatus and Systems}, vol. 103, pp. 463--473, 1984.
        
        \bibitem{Cho85}
        J.~H. Chow and P.~Kokotovi\'{c}, ``Time scale modeling of sparse dynamic
          networks,'' \emph{IEEE Transactions on Automatic Control}, vol.~30, pp.
          714--722, 1985.
        
        \bibitem{Cho91}
        R.~Date and J.~Chow, ``Aggregation properties of linearized two-time-scale
          power networks,'' \emph{IEEE Transactions on Circuits and Systems}, vol.~38,
          pp. 720--730, 1991.
        
        \bibitem{Bergen1981}
        A.~R. Bergen and D.~J. Hill, ``{A Structure Preserving Model for Power System
          Stability Analysis},'' \emph{{IEEE} Transactions on Power Apparatus and
          Systems}, vol. {PAS}-100, no.~1, pp. 25--35, jan 1981.
        
        \bibitem{Yang2018}
        Y.~Yang, J.~Zhao, H.~Liu, Q.~Z., D.~J., and Q.~J., ``A
          matrix-perturbation-theory-based optimal strategy for small-signal stability
          analysis of large-scale power grid,'' \emph{Prot. Control Mod. Power Syst.},
          vol.~3, p.~34, 2018.
        
        \bibitem{Pagnier2019b}
        L.~Pagnier and P.~Jacquod, ``{Optimal Placement of Inertia and Primary Control:
          A Matrix Perturbation Theory Approach},'' \emph{{IEEE} Access}, vol.~7, pp.
          145\,889--145\,900, 2019.
        
        \bibitem{Coletta2020}
        T.~Coletta and P.~Jacquod, ``{Performance Measures in Electric Power Networks
          Under Line Contingencies},'' \emph{{IEEE} Transactions on Control of Network
          Systems}, vol.~7, no.~1, pp. 221--231, mar 2020.
        
        \bibitem{Bamieh2020}
        \BIBentryALTinterwordspacing
        B.~Bamieh. (2020, Feb.) {A Tutorial on Matrix Perturbation Theory (using
          compact matrix notation)}. Online. [Online]. Available:
          \url{arxiv:2002.05001}
        \BIBentrySTDinterwordspacing
        
        \bibitem{J.J.Sakurai2017}
        J.~J. Sakurai and J.~Napolitano, \emph{{Modern Quantum Mechanics}},
          2nd~ed.\hskip 1em plus 0.5em minus 0.4em\relax Cambridge, U.K.: Cambridge
          University Press, 2017.
        
        \bibitem{Chow2013}
        J.~H. Chow, Ed., \emph{{Power System Coherency and Model Reduction}}.\hskip 1em
          plus 0.5em minus 0.4em\relax New York, NY, USA: Springer, 2013.
        
        \bibitem{Neumann1929}
        J.~von Neumann and E.~P. Wigner, ``{{\"U}ber das Verhalten von Eigenwerten bei
          adiabatischen Prozessen},'' \emph{Phys. Z.}, vol.~30, pp. 467--470, 1929.
        
        \bibitem{MiroslavFiedler1989}
        \BIBentryALTinterwordspacing
        M.~Fiedler, ``\BIBforeignlanguage{eng}{Laplacian of graphs and algebraic
          connectivity},'' \emph{\BIBforeignlanguage{eng}{Banach Center Publications}},
          vol.~25, no.~1, pp. 57--70, 1989. [Online]. Available:
          \url{http://eudml.org/doc/267812}
        \BIBentrySTDinterwordspacing
        
        \bibitem{Mohar1991}
        B.~Mohar, ``{The Laplacian spectrum of graphs},'' in \emph{Graph Theory,
          Combinatorics, and Applications}.\hskip 1em plus 0.5em minus 0.4em\relax
          Wiley, 1991, pp. 871--898.
        
        \bibitem{Pagnier2019}
        L.~Pagnier and P.~Jacquod, ``Inertia location and slow network modes determine
          disturbance propagation in large-scale power grids,'' \emph{{PLOS} {ONE}},
          vol.~14, no.~3, p. e0213550, mar 2019.
        
        \bibitem{Tyloo2019}
        M.~Tyloo, L.~Pagnier, and P.~Jacquod, ``The key player problem in complex
          oscillator networks and electric power grids: Resistance centralities
          identify local vulnerabilities,'' \emph{Science Advances}, vol.~5, no.~11, p.
          eaaw8359, nov 2019.
        
        \bibitem{Breulmann2000}
        H.~Breulmann, W.~Winter \emph{et~al.}, ``{Analysis and Damping of Inter-Area
          Oscillations in the UCTE / CENTREL Power System},'' CIGR{\'E}, Paris, France,
          2000.
        
        \end{thebibliography}
\end{document}